\documentclass[%
amsmath,amssymb,floatfix,
preprint,%
longbibliography,
superscriptaddress,
]{revtex4-2} 

\usepackage{dcolumn}%
\usepackage{epsfig, graphics, amssymb, amsmath, color, bm}
\usepackage{epstopdf}
\usepackage{graphicx}
\usepackage{lipsum}  

\emergencystretch=\maxdimen
\definecolor{dgreen}{rgb}{0.0, 0.7, 0.0}

\usepackage[outer=1.6cm, inner = 1.6cm, top= 2.3cm, bottom = 2.3cm]{geometry}  

\begin{document}

\title{Propulsion of a three-sphere micro-robot in a porous medium}

\author{Chih-Tang Liao}
\affiliation{Department of Mechanical Engineering,
Santa Clara University, Santa Clara, CA 95053, USA}

\author{Andrew Lemus}
\affiliation{Department of Mechanical Engineering,
Santa Clara University, Santa Clara, CA 95053, USA}

\author{Ali G\"urb\"uz}
\affiliation{Department of Mechanical Engineering,
Santa Clara University, Santa Clara, CA 95053, USA}

\author{Alan C.~H.~Tsang}
\affiliation{Department of Mechanical Engineering,
University of Hong Kong, Pokfulam Road, Hong Kong, China}

\author{On Shun Pak}\thanks{Electronic mail: opak@scu.edu}
\affiliation{Department of Mechanical Engineering,
Santa Clara University, Santa Clara, CA 95053, USA}

\author{Abdallah Daddi-Moussa-Ider}
\thanks{abdallah.daddi-moussa-ider@open.ac.uk}
\affiliation{School of Mathematics and Statistics, The Open University, Walton Hall, Milton Keynes MK7 6AA, UK}

\date{\today}
 
\begin{abstract}
Microorganisms and synthetic microswimmers often encounter complex environments consisting of  networks of obstacles embedded into viscous fluids. Such settings include biological media, such as mucus with filamentous networks, as well as environmental scenarios, including wet soil and aquifers. A fundamental question in studying their locomotion is how the impermeability of these porous media impact their propulsion performance compared with the case that in a purely viscous fluid. Previous studies showed that the additional resistance due to the embedded obstacles leads to an enhanced propulsion of different types of swimmers, including undulatory swimmers, helical swimmers, and squirmers. In this work we employ a canonical three-sphere swimmer model to probe the impact of propulsion in porous media. The Brinkman equation is utilized to model a sparse network of stationary obstacles embedded into an incompressible Newtonian liquid. We present both a far-field theory and numerical simulations to characterize the propulsion performance of the swimmer in such porous media. In contrast to enhanced propulsion observed in other swimmer models, our results reveal that both the propulsion speed and efficiency of the three-sphere swimmer are largely reduced by the impermeability of the porous medium.  We attribute the substantial reduction in propulsion performance to the screened hydrodynamic interactions among the spheres due to the more rapid spatial decays of flows in Brinkman media. These results highlight how enhanced or hindered propulsion in porous media is largely dependent on individual propulsion mechanisms. The specific example and physical insights provided here may guide the design of synthetic microswimmers for effective locomotion in porous media in their potential biological and environmental applications.
\end{abstract}


\maketitle

\newpage
\section{Introduction} \label{sec:Intro}

Fluid flows at the microscopic scale lie in the low Reynolds numbers (Re) regime, where viscous forces dominate inertial forces.  Swimming in purely viscous fluids at low Re is governed by the Stokes equation, whose time independence and kinematic reversibility impose stringent physical constraints on generating self-propulsion in the absence of inertia \cite{Purcell1977,LaugaPowers2009,Yeomans2014,PakLauga2015}. There has been substantial interest and progress in understanding the locomotion of natural and synthetic microswimmers for their fundamental biological importance as well as potential biomedical and environmental applications \cite{Fauci06,LaugaPowers2009,Nelson2010,Gao2014,Wu2020}. These microswimmers need to adopt swimming strategies different from those typically employed for inertial swimming at the macroscopic scale. For instance, Purcell's scallop theorem states that reciprocal motions (sequences of motions with time-reversal symmetry) cannot generate net self-propulsion at low Re \cite{Purcell1977}. Any reciprocal swimming strategies such as flapping motions of a rigid body employed by fish and many aquatic animals in the macroscopic world therefore would become ineffective for swimming at low Re. Extensive works have been performed on analyzing locomotion of swimming microorganisms at low Re, which have also inspired different ingenious designs of synthetic microswimmers \cite{Ebbens2010,Sengupta2012,Hu2018,Tsang2020,sharan2023pair}.

While low-Re swimming is relatively well studied in purely viscous liquids, in many real-world scenarios these microswimmers encounter complex microscopic environments consisting of irregular networks of obstacles embedded into viscous fluids \cite{Bechinger2016, Bhattacharjee2019}. Examples of these porous media include biological gels and porous tissues \cite{Leshansky2009, Thornlow2015, Datta2016, Mirbagheri2016} as well as wet soil and aquifers in environmental settings \cite{Dechesne2010, Turnbull2001, Adadevoh2016, Adadevoh2018}. The impermeability of the porous media largely influences the mobility and propulsion performance of the microswimmers. Intuitively, one might expect the additional resistance due to the embedded obstacles in the porous media to hinder the propulsion of microswimmers. In contrast, previous studies showed that the resistance from the porous media could lead to enhanced propulsion of various types of microswimmers. 
For example, experiments with \textit{C.~elegans} in saturated particulate systems \cite{Jung2010} as well as theoretical and numerical studies of undulating sheets immersed in Brinkman media \cite{Leshansky2009, Leiderman2016} found faster propulsion speeds in the porous media than in purely viscous fluids. Enhanced propulsion was also reported for squirmers \cite{Leshansky2009, nganguia_pak_2018} and some helical propellers/swimmers \cite{Leshansky2009,Ho2016}. A more recent study has examined some subtle and significant differences between torqued helical propulsion versus force-free and torque-free helical swimming in porous media \cite{Chen2020}. More complex dependence of the swimming speed on the resistance was also observed in three-dimensional flagellar swimming with emergent waveforms \cite{ho_leiderman_olson_2019}. In addition to propulsion speed, the energetic cost of moving through a medium is another important measure of locomotion performance \cite{Lighthill1975}. Previous studies showed that propulsion of different swimmers in porous media can be both faster and energetically more efficient than in purely viscous fluids \cite{Leshansky2009,nganguia_pak_2018}.

\begin{figure}[t]
    \centering
    \includegraphics[width=0.6\textwidth]{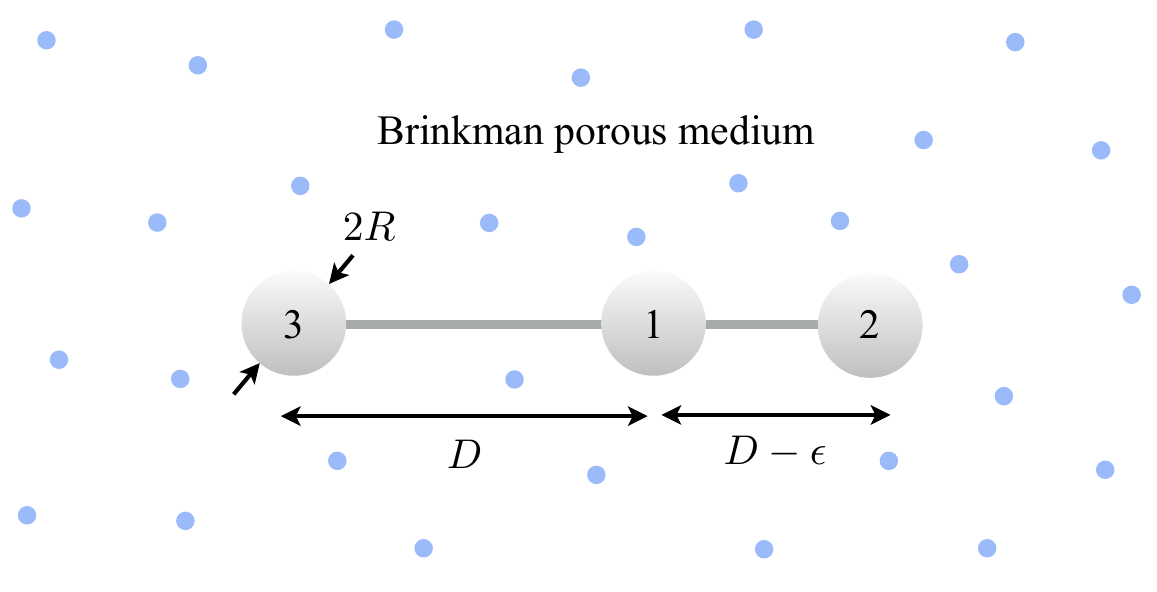}
    \caption{Schematic of the problem setup and notations. A swimmer consisting of three spheres of equal radii $R$ connected by two extensible rods immersed in a porous medium described by the Brinkman equation. The rods have a fully extended length of $D$ and a fully contracted length of $D-\epsilon$, where $\epsilon$ denotes the amount of contraction.} 
    \label{fig1}
\end{figure}

In this work, we consider the locomotion of a widely used low-Re swimmer model, a swimmer consisting of three spheres connected by two extensible rods (Fig.~\ref{fig1}), in porous media. The model was first proposed by Najafi and Golestanian \cite{Najafi2004} as a swimmer that performs non-reciprocal motions by modulating the relative distances between the spheres. The simplicity of this swimmer model has made it a useful tool for probing different aspects of low-Re locomotion \cite{Curtis2013,yasuda2023generalized,Wang2019,Nasouri2019,Pooley2007,Farzin2012,yasuda2023generalized, Zargar2009,Najafi2013,Daddi-Moussa-Ider_2018, DaddiMoussaIder2018_2,daddi2020tuning,AlanChengHouTsang2020,gurbuz2023effect}. In particular, the model has been used to examine the swimming fluctuations due to variations in local environments of heterogeneous media \cite{Jabbarzadeh2014}. Here, we utilize this three-sphere swimmer model to investigate how impermeability of porous media influences the locomotion performance in terms of both propulsion speed and efficiency. We follow previous studies on locomotion in porous media \cite{Leshansky2009, Leiderman2016, Ho2016, nganguia_pak_2018, ho_leiderman_olson_2019,Nganguia2020,CORTEZ20107609,PhysRevResearch.5.033030} in adopting an effective medium approach to model the porous medium with the Brinkman equation \cite{Brinkman1949, tam_1969, Childress1972, howells_1974,hinch_1977}. Compared with the Stokes equation for purely viscous fluids, the Brinkman equation includes the additional hydrodynamic resistance due to a sparse network of stationary obstacles embedded into the viscous flow. Numerical simulations based on Stokesian dynamics demonstrated the validity of the Brinkman equation at low volume fractions as well as its effectiveness in capturing the qualitative behaviour even for more concentrated porous media \cite{Durlofsky1987}. We will use a combination of far-field theory and finite element method numerical simulations in this work to examine the propulsion of a three-sphere swimmer in a Brinkman medium and contrast the results with previous findings for other types of swimmers.

The paper is organized as follows. In Sec.~\ref{sec:formulation}, we present the swimmer model, the governing equations, as well as the theoretical  and computational frameworks employed in this work. We then present the asymptotic results in the far-field theory in Sec.~\ref{sec:Asy}, followed by discussions on the swimming velocity and flow fields from the simulations in Sec.~\ref{sec:SwimVel} and the energetic cost of swimming in Sec.~\ref{sec:Energy}. Finally, we conclude this work with remarks on its limitations and directions for future work (Sec.~\ref{sec:conclusion}).

\section{Formulation} \label{sec:formulation}

\subsection{The swimmer model}

We examine the movement of a three-sphere swimmer within a porous medium. 
As portrayed in Fig.~1, the swimmer comprises three spheres, all with a radius of $R$, linked by two extensible rods with negligible hydrodynamic effects. The maximum length of each rod when fully extended is defined as $D$, while the minimum contracted length is $D-\epsilon$. Here, $\epsilon$ signifies the degree of contraction or extension during each stroke (henceforth referred to as the ``contraction length"). We denote the center sphere's position as $\mathbf{r}_1$, and the positions of the front and rear spheres as $\mathbf{r}_2$ and $\mathbf{r}_3$, respectively. Consequently, the length of the rods connecting adjacent spheres are defined as follows:
\begin{equation}
    \left( \mathbf{r}_2 - \mathbf{r}_1 \right) \cdot \hat{\mathbf{e}}_z = g(t) \, , \qquad
    \left( \mathbf{r}_1 - \mathbf{r}_3 \right) \cdot \hat{\mathbf{e}}_z = h(t) \, .
    \label{eq:gh}
\end{equation}
The swimmer self-propels by modulating the length of the rods, $g(t)$ and $h(t)$, in a non-reciprocal manner to break the time reversal symmetry. In this work, we adhere to the original model by Najafi and Golestanian \cite{Najafi2004}, where the rod extends or contracts at a constant speed $W$ in the four-stroke cycle  specified in Table 1: In Stroke I, the swimmer contracts its left arm, initially $D$ in length, by $\epsilon$, while maintaining the right arm's length at $D$. In Stroke II, the swimmer contracts the right arm by $\epsilon$, while the left arm remains at $D-\epsilon$. In Stroke III, the swimmer extends its left arm to achieve the fully extended $D$ length, while the right arm remains fixed at $D-\epsilon$. 
Finally, in Stroke IV, the swimmer extends its right arm to revert to the original configuration, with both arms fully extended to $D$. 
This completes one full swimming cycle, and the resulting overall displacement of the swimmer is denoted as $\Delta$.

\subsection{Governing equations}
We consider a porous medium consisting of a sparse network of stationary obstacles embedded into a viscous, incompressible Newtonian flow at low-Re, modeled by the Brinkman equation \cite{Brinkman1949},
\begin{align}
     \mu \left( \nabla^2\textbf{u} - \alpha^2 \textbf{u} \right) &=\nabla p, \label{eqn:Brinkman1} \\
     \nabla\cdot\textbf{u} &=0. \label{eqn:Brinkman2}
\end{align}
Here, $\mu$ represents the dynamic viscosity, $\alpha^2$ denotes the impermeability of the porous medium, which has the dimension of (length)$^{-2}$, while $\mathbf{u}$ and $p$ denote the fluid velocity and pressure fields, respectively. The velocity of sphere $i$ is denoted as $\mathbf{V}_i$, and the force and torque acting on it are are represented as $\mathbf{F}_i$ and $\mathbf{T}_i$, respectively, where $i=1, 2, 3$.
No-slip boundary conditions are assumed to hold on the surface of the spheres, meaning that the fluid velocity at the surface of sphere $i$ is equal to $\mathbf{V}_i$. In the absence of external forces and torques, the system is force-free,
\begin{align}
\sum_{i=1}^3 \mathbf{F}_i= \mathbf{0}, \label{eqn:force_free}
\end{align}
and torque-free,
\begin{align}
\sum_{i=1}^3 \mathbf{T}_i= \mathbf{0}. 
\end{align}
It is worth noting that the torque equilibrium condition is inherently fulfilled due to the intrinsic symmetry in the problem setup.

In the low-Re regime, the translational velocities of the three spheres are linearly related to the internal forces acting upon them given by
\begin{equation}
    \mathbf{V}_i = \frac{\mathrm{d} \mathbf{r}_i}{\mathrm{d} t} 
    = \sum_{j=1}^3 \boldsymbol{\mu}_{ij} \cdot \mathbf{F}_j \, , 
\end{equation}
where $\boldsymbol{\mu}_{ij}$ represents the hydrodynamic mobility tensor that relates the translational velocity of sphere $i$ to the force applied to sphere $j$. The hydrodynamic mobility accounts for the influence of multiple fluid-mediated interactions between suspended particles. In this context, we simplify matters by focusing solely on self-interactions $(i=j)$ and pairwise interactions $(i \ne j)$ for the hydrodynamic effects. We will perform the far-field analysis in Sec. \ref{sec:Far-field} and evaluate the accuracy of the approach through a direct comparison with fully resolved numerical simulations based on the finite element method in Sec. \ref{sec:FEM}.

\begin{table}[]
\renewcommand{\arraystretch}{1.5}
    \centering
    \begin{tabular}{|c|c|c|c|c|c|}
    \hline
     ~~Stroke~~ & ~~Time interval~~ & $g(t)$ & $h(t)$ & ~~$\dot{g}(t)$~~ & ~~$\dot{h}(t)$~~ \\
     \hline\hline
     I & $[0,T/4]$ & $D$ & $D-Wt$ & 0 & $-W$ \\
     II & $[T/4,T/2]$ & ~~$D+\epsilon-Wt$~~ & $D-\epsilon$ & $-W$ & 0 \\
     III & $[T/2, 3T/4]$ & $D-\epsilon$ & $D-3\epsilon+Wt$ & 0 & $W$ \\
     IV & $[3T/4,T]$ & $D-4\epsilon+Wt$ & $D$ & $W$ & 0 \\
     \hline
    \end{tabular}
    \caption{The swimmer undergoes a four-stroke cycle specified by the temporal variation of the distance between spheres 1 and 2, $g(t)$, and that between spheres 1 and 3, $h(t)$.}
    \label{tab:my_label}
\end{table}

\subsection{Far-field theory} \label{sec:Far-field}
We consider a far-field theory to examine the scenario where the spheres are sufficiently far apart from each other. In a Brinkman medium, the hydrodynamic self mobility function is given by
\begin{equation}
    \mu_\mathrm{S} = \frac{1}{6\pi\eta R A} \, , 
    \label{eq:self-mobi-Brinkman}
\end{equation}
where $ A = 1 + \alpha R + \left( \alpha R \right)^2/9$. The pair mobility function for axial motion along the line connecting the centers of two spheres separated a distance~$d$ apart can conveniently be approximated as~\cite{felderhof2019retarded}
\begin{equation}
    \mu_\mathrm{P} = \frac{B^2 - \left( 1+\alpha d\right) e^{\alpha \left( 2R-d\right)}}{2\pi\eta d^3 \alpha^2 A^2} \, ,     \label{eq:pair-mobi-Brinkman}
\end{equation}
where $B = 1 + \alpha R + \left( \alpha R \right)^2/3$. 
It is worth noting that several attempts have been made in the past to approximate pair mobility functions, but few of these have resulted in accurate predictions~\cite{bonet1991dynamics, ardekani2006unsteady, tatsumi2013propagation}.
The Stokes regime is attained in the limit $\alpha \to 0$ as $\mu_\mathrm{P} \rightarrow \frac{3d^2-2R^2}{12\pi\eta d^3}$, 
indicative of the classical expression derived from the Rotne-Prager tensor~\cite{wajnryb2013generalization}.

Taking the derivative of Eq.~\eqref{eq:gh} with respect to time leads to the relationships $V_2 = V_1 + \dot{g}$ and $V_3 = V_1 - \dot{h}$, where the dots signify a time derivative, and $V_i =  \mathbf{V}_i \cdot \hat{\mathbf{e}}_z$, where $\hat{\mathbf{e}}_z$ represents the unit vector along the axial ($z$--) direction (Fig.~1). Enforcing the force-free condition, Eq.~\eqref{eqn:force_free}, the instantaneous axial velocity of the center sphere is determined as
\begin{equation}
    V_1 = \frac{1}{K} \left( \left( \mu_\mathrm{S} -\mu_{12} \right) \dot{h} M_+ - \left( \mu_\mathrm{S} -\mu_{13} \right) \dot{g} M_- \right) \, , 
    \label{eq:V1_final-Brinkman}
\end{equation}
where we have defined, for convenience, the abbreviations 
\begin{align}
M_\pm = \mu_\mathrm{S} \pm \mu_{12} \mp \mu_{13} - \mu_{23},
\end{align}
and
\begin{align}
K = 3\mu_\mathrm{S}^2 - 2\Lambda \mu_\mathrm{S} - N,
\end{align}
with $ \Lambda = \mu_{12} + \mu_{13} + \mu_{23}$ and $N = \mu_{12}^2 + \left( \mu_{13} - \mu_{23} \right)^2 - 2 \mu_{12} \left( \mu_{13} + \mu_{23} \right)$.
It is worth recalling that the pair mobilities are computed as $\mu_{12} = \mu_\mathrm{P} \left( d=g \right) $, $\mu_{13} = \mu_\mathrm{P} \left( d=h \right) $, and $\mu_{23} = \mu_\mathrm{P} \left( d=g+h \right)$. Ultimately, the average swimming velocity is computed by taking the mean over a complete swimming cycle as
\begin{equation}
    \overline{V}_1 = \frac{1}{T} \int_0^T V_1(t) \, \mathrm{d} t \, . 
    \label{eq:mean_velocity}
\end{equation}
A full analytical evaluation of the average swimming speed based on Eq.~\eqref{eq:V1_final-Brinkman} is intractable even in the simplest scenario of Stokes flows~\cite{gurbuz2023effect}. We therefore calculate the average swimming speed by evaluating the integral in Eq.~\eqref{eq:mean_velocity} numerically using the mobilities in the far-field theory. 

\subsection{Finite element method simulations} \label{sec:FEM}

To fully capture the hydrodynamic interaction between the spheres, we solve the Brinkman equation, Eq.~\eqref{eqn:Brinkman1}, and continuity equation, Eq.~\eqref{eqn:Brinkman2} using the finite element method (FEM) conducted in the COMSOL Multiphysics environment. We utilize the creeping flow interface with an added volume force to account for the resistance term, $-\mu \alpha^2 \mathbf{u}$, in the Brinkman equation. Similar to the Stokes equation, the time independence of the Brinkman equation allows the simulation of the swimming motion to be divided into a series of stationary simulations at different time instants over a swimming cycle. For each stationary simulation, we solve the momentum and continuity equations together with the force-free condition, Eq.~\eqref{eqn:force_free}, implemented as a global equation, to obtain the instantaneous swimming speed, velocity and pressure fields simultaneously at each time instant. The total net displacement of the swimmer and average swimming speed are then determined through numerical integration over the full swimming cycle.

The simulation is set up as a two-dimensional axisymmetric domain where the three spheres are aligned along the axis of symmetry. To mitigate the effect of confinement, the domain extends $1000R$ in each direction from the outer spheres, and the radius of the confining cylinder is maintained at $1000R$. Each simulation comprises of approximately 22,000 P3-P2 (third-order for fluid velocity and second-order for pressure) triangular mesh elements, with local mesh refinement applied around the spheres. The Multifrontal Massively Parallel Sparse (MUMPS) direct solver is used for all simulations. In addition to comparing with predictions from the far-field theory, we validated our numerical approach against analytical results of the drag acting on a translating sphere in a Brinkman medium \cite{howells_1974} and propulsion of the three-sphere swimmer in the Stokes regime \cite{Najafi2004}.

\subsection{Non-dimensionalization}
Hereafter we will scale lengths by $R$, times by $T$, and forces by $\eta R^2/T$. Consequently, velocities are scaled by $R/T$ and powers are scaled by $\eta R^3 /T^2$. Furthermore, we define the dimensionless numbers $\delta = \epsilon/D$, $\sigma = R/D$, and~$\lambda = \alpha D$. Henceforth, we refer to only dimensionless variables and use the same symbols as their corresponding dimensional counterparts for convenience.

\section{Results and Discussion}
\subsection{Asymptotic results} \label{sec:Asy}
In the far-field theory, we can make analytical progress by considering the asymptotic limit of $\sigma \ll 1$ and expanding the velocity perturbatively in terms of the small parameter $\sigma$. By substituting the expressions for the self- and pair-mobility functions, as given in Eqs.~\eqref{eq:self-mobi-Brinkman} and~\eqref{eq:pair-mobi-Brinkman}, into Eq.~\eqref{eq:V1_final-Brinkman}, and recognizing that $N$ is of the order $\mathcal{O}\left( \sigma^2 \right)$, we can readily determine the instantaneous swimming velocity through a Taylor expansion with respect to the small parameter $\sigma$ as
\begin{equation}
    \overline{V}_1 = \Psi + \frac{\Gamma +  \Phi}{\lambda^2} \, , \label{eq:v1}
\end{equation}
where we have defined
\begin{align}
    \Psi &= \frac{1}{3} \left( 2 \operatorname{E}_1 \left( \lambda ( 2-\delta ) \right) - \operatorname{E}_1 \left( 2\lambda (1-\delta) \right) - \operatorname{E}_1 (2\lambda)\right),\\ 
   \Gamma &= \frac{\delta^2 \left( 186-372\delta+297\delta^2-111\delta^3+16\delta^4 \right)}{12  \left( 1-\delta \right)^3 \left( 2-\delta \right)^2},
    \label{eq:Gamma}\\
\Phi &= \beta_1 e^{-\lambda} + \beta_2 e^{-2\lambda} + \beta_3 e^{-\lambda (1-\delta)} + \beta_4 e^{-\lambda (2-\delta)} + \beta_5 e^{-2\lambda (1-\delta)} \, , 
\end{align}
wherein $\operatorname{E}_1$ is the En-function with $n=1$, related to the exponential integral $\operatorname{Ei}$ via $\operatorname{E}_1 (x) = -\operatorname{Ei}(-x)$, for a positive real number~$x$, and  
\begin{subequations}
    \begin{align}
    \beta_1 &= \frac{4\delta (1+\lambda)}{3} \, , \\
    \beta_2 &= \frac{1+2\lambda}{12} \, , \\
    \beta_3 &= -\frac{4\delta \left( 1+\lambda (1-\delta) \right)}{3 \left( 1-\delta \right)^3} \, , \\
    \beta_4 &= -\frac{2 \left( 1+ \lambda (2-\delta) \right)}{3 \left( 2-\delta \right)^2} \, , \\
    \beta_5 &= \frac{1 + 2\lambda (1-\delta)}{12 \left( 1-\delta \right)^2} \, .
\end{align}
\end{subequations}
In the Stokes limit ($\lambda \to 0$), the mean swimming velocity reduces to 
\begin{equation}
    \overline{V}_1 = \frac{1}{3} \left( \frac{2\delta^2}{1-\delta} 
    + \ln \left( \frac{4(1-\delta)}{\left( 2-\delta \right)^2} \right) \right) \, ,
\end{equation}
in full agreement with the results derived in \cite{gurbuz2023effect}.

\begin{figure*}[t]
    \centering
    \includegraphics[width=0.95\textwidth]{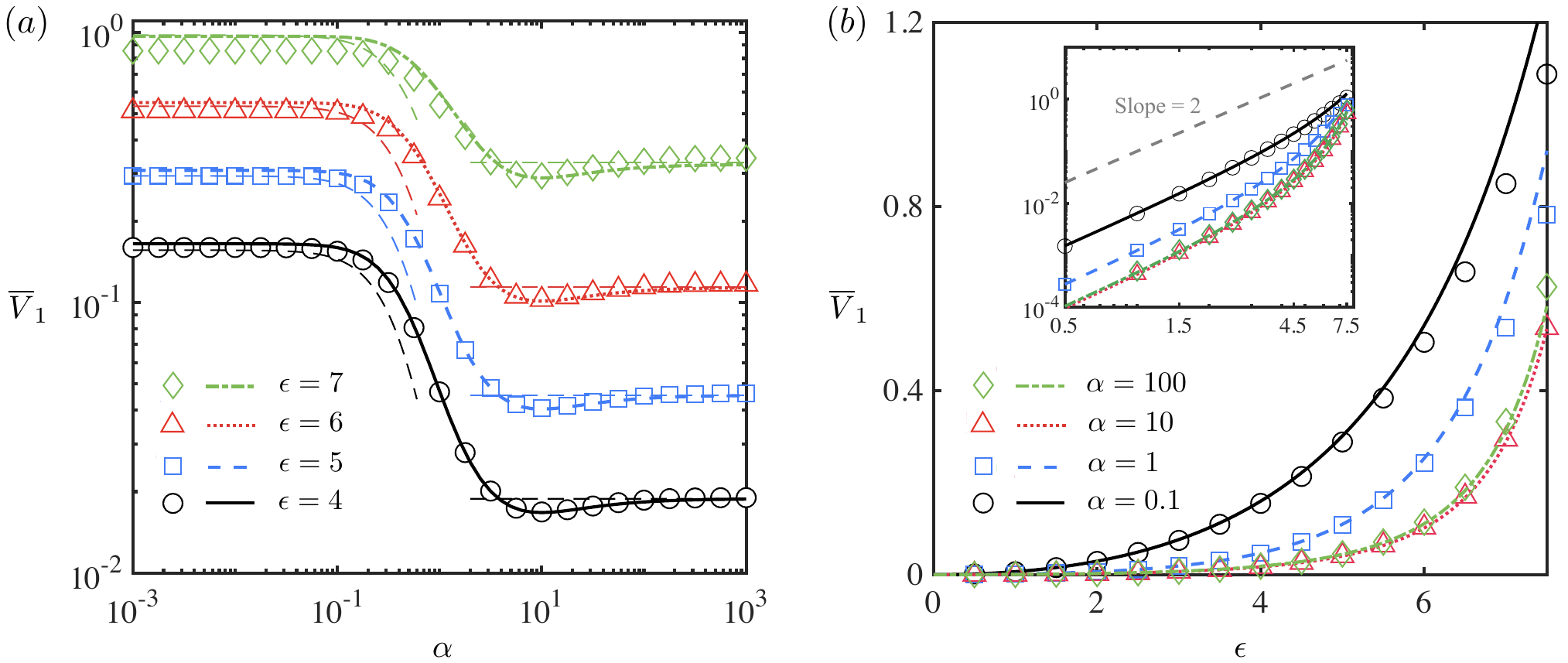}
    \caption{Swimming velocity in a Brinkman medium. (a) The mean swimming velocity, $\overline{V}_1$, as a function of the resistance, $\alpha$, for different values of contraction length, $\epsilon$ (see legends). The lines correspond to predictions based on the far-field theory (Eqs.~\eqref{eq:V1_final-Brinkman} and  \eqref{eq:mean_velocity}), which agree well with results obtained by FEM simulations (symbols). The thin dotted lines on the left and right of the panel represent, respectively, asymptotic results given by Eq.~\eqref{eq:v1} and Eq.~\eqref{eqn:LargeLambda}. (b) The mean swimming velocity, $\overline{V}_1$, as a function of the contraction length, $\epsilon$, for different values of resistance, $\alpha$ (see legends). Results based on the far-field theory (lines) agree well with those by FEM simulations (symbols). The inset displays a log-log plot of the results, with a dotted gray line of slope 2 added for comparison. In all results, we set $D=10$.} 
    \label{fig:MeanVel}
\end{figure*}

By conducting a Taylor expansion of Eq.~\eqref{eq:v1} with respect to $\delta = 0$, we can approximate the mean swimming speed as
\begin{equation}
    \overline{V}_1 = \left( \frac{\delta}{\lambda} \right)^2 \left( X_0 - X_1 e^{-\lambda}
    + X_2 e^{-2\lambda} \right) , 
\end{equation}
where $X_0 = (31/8) \left( 1+2\delta \right)$, 
$X_1 = c_1 + c_2 \delta $, and $X_2 = c_3 + c_4 \delta $, with
\begin{subequations}
    \begin{align}
    c_1 &= \frac{4}{3} \left( 3+3\lambda+\lambda^2 \right) \, , \\
    c_2 &= \frac{2}{3} \left( 12+12\lambda+5\lambda^2+\lambda^3 \right) \, , \\
    c_3 &= \frac{1}{24} \left( 3+6\lambda+4\lambda^2 \right) \, , \\
    c_4 &= \frac{1}{12} \left( 3+6\lambda+5\lambda^2+2\lambda^3 \right) \, .
\end{align}
\end{subequations}
It can readily be shown that in the limits $\lambda \to 0$ and $\delta \to 0$ that $\overline{V}_1 = 7\delta^2 \left( 1+\delta \right)/12$.

We also examine the limit of large resistance ($\lambda \to \infty$) in the Brinkman medium. In this regime, the swimming velocity  can likewise be Taylor expanded with respect to the parameter $\sigma$ to obtain the result
\begin{equation}
    \overline{V}_1 = \sigma^2 \Gamma, \label{eqn:LargeLambda}
\end{equation}
where $\Gamma$ is given by Eq.~\eqref{eq:Gamma}. It is interesting to note that the far-field theory predicts a finite mean swimming speed in the limit of infinite resistance of the porous medium, a result we verify by numerical simulations in later sections. Furthermore, in the limit $\lambda \to \infty$ and for $\delta \to 0$, it can be verified that $\overline{V}_1 = 31\left( \delta \sigma \right)^2 \left( 1+2\delta \right)/8$.

\subsection{Swimming velocity and flow field} \label{sec:SwimVel}
In this section, we discuss the swimming velocity and compare predictions based on the far-field theory (Sec.~\ref{sec:Far-field}), including asymptotic results given in Sec.~\ref{sec:Asy}, with results obtained by numerical simulations via FEM described in Sec.~\ref{sec:FEM}. 

Fig.~\ref{fig:MeanVel}(a) displays the variation of the mean swimming velocity, $\overline{V}_1$, as a function of the resistance, $\alpha$, of the porous medium. As the resistance of the porous medium increases from the Stokes limit ($\alpha=0$), the swimmer begins to experience a significant speed reduction when $\alpha$ is in the range of $\mathcal{O}(0.1)-\mathcal{O}(1)$. This is in stark contrast to enhanced propulsion speeds in porous media observed for various types of swimmers reported in previous studies \cite{Leshansky2009, Jung2010,Leiderman2016, Ho2016,nganguia_pak_2018, ho_leiderman_olson_2019}, including undulatory swimmers, helical swimmers, and squirmers. The qualitatively different behavior revealed here highlights whether resistance in a porous medium enhances or hinders propulsion could largely depend on individual propulsion mechanisms. The swimmer continues to slow down as $\alpha$ increases, reaching a local minimum when $\alpha \approx \mathcal{O}(10)$. Interestingly, a further increase in $\alpha$ then leads to a slight speed enhancement, plateauing a finite swimming speed for very large resistance as predicted by the asymptotic result, Eq.~\eqref{eqn:LargeLambda} [thin dotted lines in Fig.~\ref{fig:MeanVel}(a)]. The same general characteristics are observed for different values of contraction lengths, $\epsilon$, in both results by the far-field theory (lines) and FEM simulations (symbols), which agree well as shown in Fig.~\ref{fig:MeanVel}(a). Predictions by the far-field theory deviate more from results by FEM simulations for larger values of contraction length, $\epsilon$, as expected, where the spheres become in closer proximity to each other.

In the Stokes limit ($\alpha=0$), it was known that the swimming velocity increases quadratically with $\epsilon$ \cite{Najafi2004,Earl2007}. More recently, it has been shown that the presence of axisymmetric confinement modifies the quadratic dependence of the swimming velocity on $\epsilon$ \cite{gurbuz2023effect}. Here, we examine how the swimming velocity varies with $\epsilon$ in porous media with different values of resistance. As shown in Fig.~\ref{fig:MeanVel}(b), although the presence of resistance in porous media slows down the swimmer, the swimmer retains the quadratic dependence with $\epsilon$; the inset displays the results in log-log scale, demonstrating that the curves attain slopes of approximately two for small values of $\epsilon$, in contrast to the case of the confined three-sphere swimmer \cite{gurbuz2023effect}.

\begin{figure*}[t]
    \centering
    \includegraphics[width=1\textwidth]{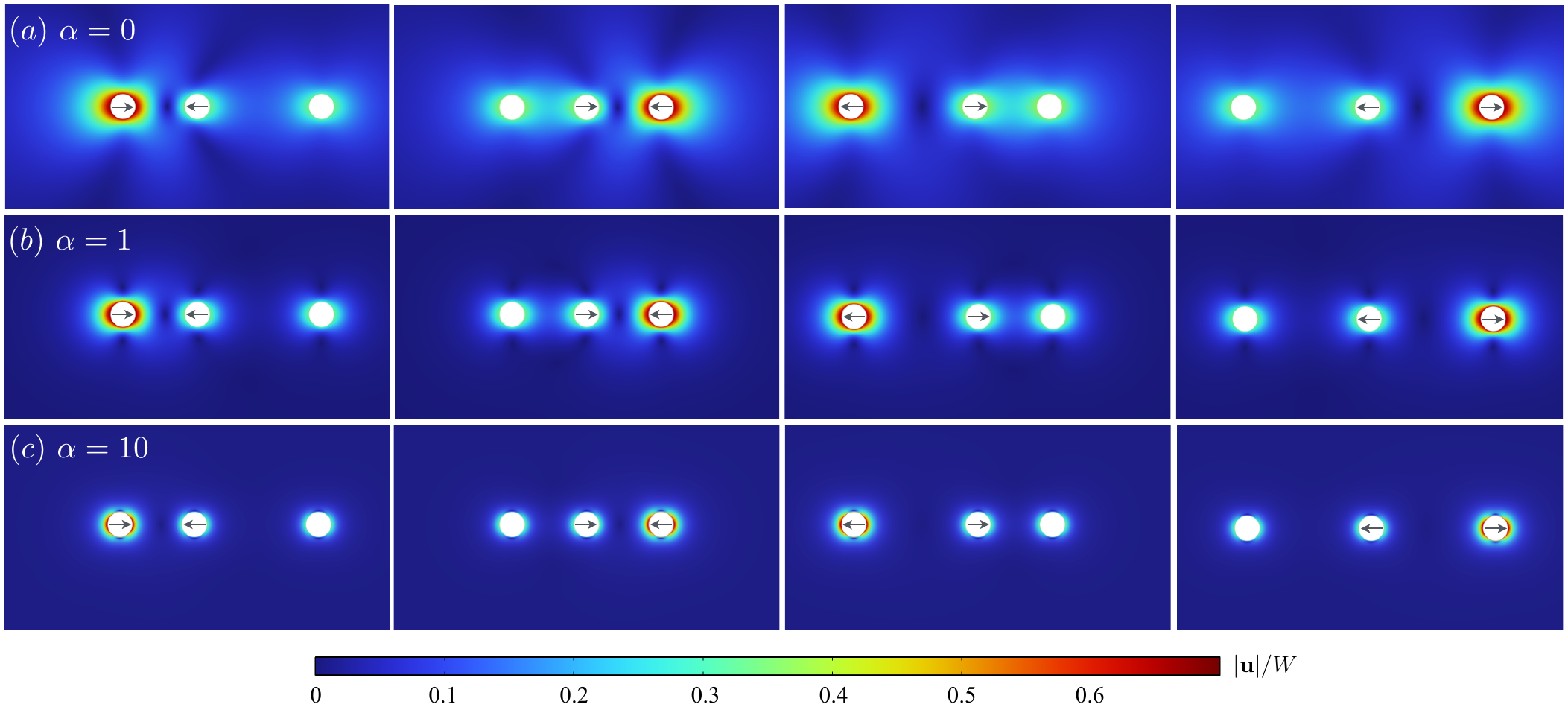}
    \caption{The flow speed distribution around the three-sphere swimmer at the end of strokes I to IV (from left to right) for different values of dimensionless resistance of the Brinkman medium: (a) $\alpha =0$, (b) $\alpha =1$, and (c) $\alpha=10$. Here the colormap displays the magnitude of the dimensional flow speed $|\mathbf{u}|$ scaled by the constant contraction/extension speed $W$ of the rods. The arrows indicate the relative motion of the pair of spheres in different strokes. Here, $D=10$ and $\epsilon=4$.}
    \label{fig:FlowField}
\end{figure*}

To gain some physical insights into the observed reduction in propulsion speed of a three-sphere swimmer in porous media, we reiterate that the propulsion mechanism of the swimmer relies on the hydrodynamic interaction between the spheres. Specifically, the interactions between the spheres via their surrounding flows lead to only partial cancellation of the displacements generated by stokes I and II, giving rise to the net displacement of the swimmer (strokes III and IV are related to strokes I and II by symmetries; see \cite{Najafi2004,gurbuz2023effect} for more detailed discussions). In a purely viscous fluid, a Stokeslet decays as $1/r$ spatially with the distance $r$ away from the singularity, whereas a Brinkmanlet decays much faster as $1/r^3$ in a porous medium due to the screening of velocity disturbance \cite{koch_brady_1985, Koch1998, Long2001}. The more rapid spatial flow decay therefore weakens the hydrodynamic interactions between spheres in porous media, hindering the propulsion of the three-sphere swimmer. To visualize the effect, we display the flow field around a three-sphere swimmer performing different strokes during a swimming cycle in Fig.~\ref{fig:FlowField}. The flow fields in the Stokes limit ($\alpha = 0$) are displayed as benchmarks for comparison (Fig.~\ref{fig:FlowField}a). As the resistance increases [$\alpha = 1$ in (b) and $\alpha=10$ in (c) ], the flow around individual spheres is observed to decay more rapidly away from the spheres as a result of the hydrodynamic screening effect in Brinkman media. The aforementioned physical picture resembles the mechanism by which the propulsion speed of a three-sphere swimmer undergoes reduction due to the confinement effect \cite{gurbuz2023effect}: the presence of confinement also induces more rapid spatial decays of the flow of moving bodies, thereby weakening the hydrodynamic interactions between spheres and hence propulsion speed of the swimmer inside a tube.

\subsection{Power dissipation and swimming efficiency} \label{sec:Energy}

In addition to propulsion speed, the energetic cost expended by the swimmer during the swimming motion is another important property of locomotion. In this section, we examine how resistance influences the energetic cost of swimming through a porous medium.

\begin{figure*}[t]
    \centering
    \includegraphics[width=1\textwidth]{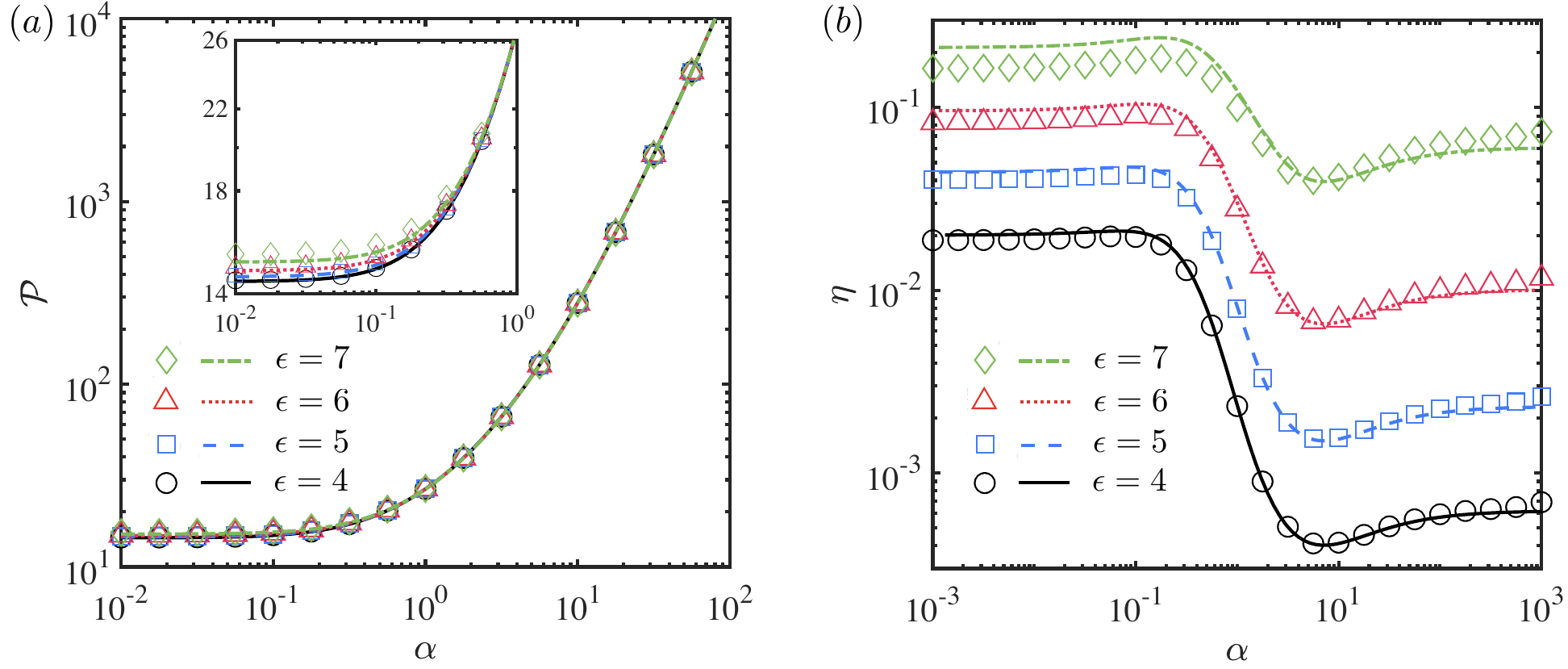}
    \caption{Schematic of the problem setup and notations. (a) A swimmer consisting of three spheres of equal radii $R$ connected by two extensible rods immersed in a porous medium described by the Brinkman equation. The rods have a fully extended length of $D$ and a fully contracted length of $D-\epsilon$, where $\epsilon$ denotes the amount of contraction. (b) The swimmer undergoes a four-stroke cycle designed by Najafi and Golestanian \cite{Najafi2004} to produce a net displacement, $\Delta$.
    Solid curves depict the mean power, computed through the far-field theory outlined in Eq.~\eqref{eq:mean_power}, and the efficiency, as defined by Eq.~\eqref{eq:lighthill_def}. The efficiency is calculated from the results of the far-field theory, where the actual and towing powers are determined by Eqs.~\eqref{eq:mean_power} and \eqref{eq:P_two}, respectively.
    } 
    \label{fig:power}
\end{figure*}

The instantaneous power dissipation by the swimmer, $P$, is given by the cumulative sum of power dissipated by each translatin sphere as
\begin{equation}
    P = \sum_{i=1}^3 \mathbf{F}_i \cdot \mathbf{V}_i \, .
\end{equation}
Because of the axisymmetric nature of the problem and the swimmer being force-free, the power dissipation can be succinctly expressed as~\cite{GolestanianAjdari2008}
\begin{equation}
    P = \dot{g} F_2 - \dot{h} F_3 \, .
\end{equation}
Upon substituting the expressions for the forces and actuation rates, the instantaneous power can be conveniently cast in the form
\begin{equation}
        P = \frac{2}{K} \bigg( \left( \mu_\mathrm{S}-\mu_{13} \right) \dot{g}^2 
    + H \, \dot{g} \dot{h}
    + \left( \mu_\mathrm{S}-\mu_{12} \right) \dot{h}^2
      \bigg) \, ,
      \label{eq:power_instant}
\end{equation}
where we have defined $H = \mu_\mathrm{S} - \mu_{12}-\mu_{13}+\mu_{23}$. We calculate the mean power dissipation,
\begin{align}
\overline{P} = \frac{1}{T} \int_0^T P(t) \, \text{d}t, \label{eq:mean_power}
\end{align}
by averaging Eq.~\eqref{eq:power_instant} over a full swimming cycle. In the main text, we perform the integral in Eq.~\eqref{eq:mean_power} numerically using the mobilities in the far-field theory. Fig.~\ref{fig:power}(a) shows that the power dissipation by the swimmer grows substantially with the resistance in the porous medium. As the swimmer's rods contract more (increased $\epsilon$), there is a corresponding elevation in power expenditure during the swimming motion. However, this dependency is less conspicuous; see inset for a magnified view. As a remark, similar to the treatment of swimming velocity in Sec.~\ref{sec:Asy}, analytical expressions of the power dissipation can be derived in various asymptotic limits, which we consider in the Appendix \ref{sec:Appendix}.

As a measure of the swimming efficiency, Lighthill introduced the Froude efficiency, a concept coming from propeller theory, defined as \cite{Lighthill1952,Lighthill1975}
\begin{align}
\eta = \frac{\overline{P}}{\overline{P}_\text{tow}}, \label{eq:lighthill_def}
\end{align}
which compares power dissipation during swimming motion, $\overline{P}$, with the power required to tow the swimmer at the same average swimming speed, $\overline{P}_\text{tow}$. The definition of the towing power becomes ambiguous when applied to swimmers experiencing substantial body deformations, such as the three-sphere swimmer under consideration. We follow the approach in Nasouri \textit{et al.} \cite{Nasouri2019} to define $\overline{P}_\text{tow}$ as  the power required to tow a spherical cargo with the same total volume as the three spheres of the swimmer at the average swimming speed, $\overline{V}_1$, in a Brinkman medium, 
\begin{equation}
    \overline{P}_\text{tow} = 6\pi R_\text{eff}
    \left( 1 + \alpha R_\text{eff} + \frac{1}{9} \left( \alpha R_\text{eff}\right)^2 \right) \overline{V}_1^2 \, , 
    \label{eq:P_two}
\end{equation}
where the effective radius of the spherical cargo (scaled by $R$) is given by $R_\text{eff} = \sqrt[3]{3}$. 

We calculate the swimming efficiency based on the above definition and display the results in Fig.~\ref{fig:power}(b). Generally, the variation of the swimming efficiency with resistance in the porous medium exhibits similar characteristics to the observed variation in propulsion speed with resistance: the efficiency undegoes a significant reduction when $\alpha$ falls in the range of $\mathcal{O}(0.1)-\mathcal{O}(1)$, reaching a local minimum around $\mathcal{O}(10)$ before slowly increasing to a finite value for large resistance. In the limit of $\lambda \rightarrow \infty$, one can derive asympotically an analytical expression for the swimming efficiency as $\eta = 9\sigma^6 \Gamma^2/(32\delta^2)$, which can be further approximated as $\eta = 8649\delta^2 \sigma^6/2048$ in the regime of $\delta \ll 1$. 

Taken together, the results on propulsion speed (Fig.~\ref{fig:MeanVel}) and swimming efficiency (Fig.~\ref{fig:power}) show that the propulsion of a three-sphere swimmer in porous media is both largely slower and less efficient compared with that in a purely viscous fluid.

\section{Concluding remarks}\label{sec:conclusion}

In this work, we examine the propulsion of a three-sphere swimmer in porous media modeled by the Brinkman equation. We present results based on a far-field theory and numerical simulations via finite element method, which display satisfactory agreements especially when the spheres are sufficiently far apart from each other. Previous studies have shown that the additional resistance in porous media leads to enhanced propulsion of different types of swimmers, including undulatory swimmers, helical swimmers, and squirmers \cite{Leshansky2009, Jung2010,Leiderman2016, Ho2016,nganguia_pak_2018, ho_leiderman_olson_2019}. In stark contrast, here we demonstrate a specific example of a swimmer whose propulsion performance is substantially hindered in porous media. These results highlight whether the resistance of porous media enhances or degrade propulsion can largely depend on the specific swimming mechanisms. For instance, for swimmers that propel based on the anisotropic nature of drag acting on their slender bodies such as helical swimmers, the additional resistance enhances the drag anisotropy and thus the propulsion performance in porous media \cite{Leshansky2009,Jung2010,Chen2020}. On the other hand, the propulsion of a three-sphere swimmer considered here hinges on the hydrodynamic interaction among the moving spheres. In this case, the additional resistance due to the network of obstacles acts to screen the hydrodynamic interactions between the spheres, causing the substantial reduction in propulsion speed in porous media. Furthermore, we show that its propulsion is  energetically less efficient in porous media compared with swimming in a purely viscous fluid. 

We remark on several limitations of the current work and suggest possible directions for future research. First, we model the porous medium surrounding the swimmer by the Brinkman equation in this work. Such an effective medium approach was shown accurate in describing flows in porous media at low volume fractions \cite{Durlofsky1987}. For larger volume fractions, while it still captures qualitatively the behaviors in moderately concentrated porous media, the Brinkman equation experiences a diminishing quantitative predictive accuracy. The results in this work for large values of resistance therefore may only hold qualitatively. This limitation calls for numerical studies capable of accurately describe the dynamics in concentrated porous media in future studies \cite{Durlofsky1987,Leshansky2009}. 
Second, here we treat the embedded obstacles in the porous medium as a stationary network. It would be interesting to investigate how deformability of the network impact the propulsion performance of the three-sphere swimmer \cite{Fu_2010,wróbel_lynch_barrett_fauci_cortez_2016,moradi_shi_nazockdast_2022,Schuech2022}. Finally, we consider here swimming gaits originally proposed by Najafi and Golestanian \cite{Najafi2004} when the three-sphere swimmer was first introduced. Recent research has delved into the optimization of these gaits in the Stokes regime \cite{Alouges2008, Alouges2009, Felderhof2015,Wang2019,Nasouri2019}.  An investigation is currently underway to examine how impermeability of porous media influences the optimality of these swimming gaits and will be reported in a future work.

\section*{Acknowledgements}
O.S.P acknowledges funding support by the National Science Foundation (Grant Numbers 1931292 and 2323046). 
We are also grateful for the computational resources from the WAVE computing facility (enabled by the E.~L.~Wiegand Foundation) at Santa Clara University.\\

\appendix 

\section{Asymptotic results on power dissipation} \label{sec:Appendix}

Similar to the asymptotic analysis of swimming velocity in Sec.~\ref{sec:Asy}, by considering the limit of small $\sigma$, the mean power dissipation of the swimmer can be obtained as
\begin{equation}
    \overline{P} =  64\pi \left(1+\lambda \right) \delta^2 
    + \left( \Psi_\mathrm{P} + \frac{\Gamma_\mathrm{P}  + \Phi_\mathrm{P}}{\lambda^2} \right) \sigma , 
\end{equation}
where we have defined
\begin{align}
        \Psi_\mathrm{P} &=
    32\pi \delta \big( \operatorname{E}_1\left( 2\lambda (1-\delta) \right) - \operatorname{E}_1\left( 2\lambda \right) + 2 \left( \operatorname{E}_1\left( \lambda (1-\delta) \right) - \operatorname{E}_1\left( \lambda \right)  \right)
    \big), \\
     \Gamma_\mathrm{P} &= 8\pi \,
    \frac{\delta^2 \left( 2-\delta\right) \left( 5-5\delta-4\delta^2 \right)}{\left( 1-\delta \right)^3} \, ,\\
    \Phi_\mathrm{P} &= \xi_1 e^{-\lambda} + \xi_2 e^{-2\lambda}
    + \xi_3 e^{-\lambda \left( 1-\delta\right)}
    + \xi_4 e^{-2\lambda \left( 1-\delta\right)} \, ,
\end{align}
with 
\begin{subequations}
    \begin{align}
    \xi_1 &= 32\pi \delta \left( 1+\lambda\right) \left( 2+\delta \right)  \, , \\
    \xi_2 &= 8\pi \delta \left( 1+2\lambda \right)  \, , \\
    \xi_3 &= -\frac{32\pi \delta}{\left( 1-\delta\right)^3} 
    \left( 2-3\delta\right) \left( 1 + \lambda \left( 1-\delta\right) \right) \, , \\
    \xi_4 &= -\frac{8\pi \delta}{\left( 1-\delta\right)^2} \left( 1 + 2\lambda \left( 1-\delta \right) \right) \, .
\end{align}
\end{subequations}
In the limit $\lambda \to 0$, which corresponds to swimming in the Stokes  regime, the mean power dissipation reduces to
\begin{equation}
    \overline{P} = 16\pi \delta
    \left( 4\delta -  \sigma \left( \frac{\delta (2-\delta)}{1-\delta} + 6\ln \left( 1-\delta \right) \right)  \right) .
    \label{eq:power_stokes}
\end{equation}
Performing a Taylor expansion of Eq.~\eqref{eq:power_stokes} around $\delta = 0$ yields
\begin{equation}
    \overline{P} = \left( \frac{\delta}{\lambda} \right)^2 \left( Z_0 - \sigma \left( Z_1 e^{-\lambda}
    + Z_2 e^{-2\lambda} \right) \right) , 
\end{equation}
where 
\begin{align}
    Z_0 &= 8\pi \Big( 8\lambda^2 +  \left( 2 \left( 5+4\lambda^3\right)+15\delta \right) \sigma \Big) \, ,\\
    Z_1 &= d_1 + d_2 \delta, \\
    Z_2 &= d_3 + d_4 \delta,
 \end{align}   
 and
\begin{align}
    d_1 &= 64\pi \left( 1+\lambda \right) \, , \\
    d_2 &= 32\pi \left( 3+3\lambda+\lambda^2 \right) \, , \\
    d_3 &= 16\pi \left( 1+2\lambda \right) \, , \\
    d_4 &= 8\pi \left( 3+6\lambda+4\lambda^2 \right) \, .
\end{align}
Furthermore, in the limits $\lambda \to 0$ and $\delta \to 0$, we obtain
\begin{equation}
    \overline{P} = 32\pi \delta^2 \left( 2 + (2+\delta) \sigma \right) \, .
\end{equation}
By investigating the scenario when $\lambda \to \infty$, it becomes apparent that the dissipated power experiences a quadratic growth with the dimensionless impermeability coefficient. To leading order, it can be evaluated as
\begin{equation}
    \overline{P} = \frac{64}{9} \, \pi \left( \delta \sigma \lambda \right)^2  \, .
\end{equation}

Similarly, analytical results for the mean towing power, $\overline{P}_\text{tow}$, can be obtained by considering different asymptotic limits. To leading order in $\sigma$, the mean towing power can be obtained by inserting the expression of the averaged swimming speed given by Eq.~\eqref{eq:v1} into Eq.~\eqref{eq:P_two}. In the limit as $\lambda \to 0$, corresponding to the Stokes regime, the mean towing power can be presented in a scaled form as
\begin{equation}
    \overline{P}_\text{tow} = \frac{2\pi}{3^\frac{2}{3}} \, \sigma^2 
    \left( \frac{2\delta^2}{1-\delta} 
    + \ln \left( \frac{4(1-\delta)}{\left( 2-\delta \right)^2} \right) \right)^2.
\end{equation}
Considering next the limit $\lambda \to \infty$, it can readily be shown that the mean towing power also scales quadratically with~$\lambda$ as
\begin{equation}
    \overline{P}_\text{tow} = 2\pi \left( \sigma^4 \lambda \Gamma \right)^2 \, , 
\end{equation}
which for $\delta \to 0$ is obtained as
\begin{equation}
    \overline{P}_\text{tow}  = \frac{961}{32} \, \pi \left( \sigma^4 \delta^2 \lambda \right)^2 \, .
\end{equation}

%


\end{document}